\documentclass[aoas,MSNbibl,nameyear,seceqn,dvips]{arximspdf}
\usepackage{dcolumn}
\usepackage{graphicx}

\doi{10.1214/13-AOAS639} 
\volume{7}
\issue{2}
\pubyear{2013}
\firstpage{691}
\lastpage{713}

\makeatletter

\newcolumntype{d}[1]{D{.}{.}{#1}}

\newcommand{\en}{\enskip}

\renewcommand{\middle}{}

\newcommand{\lp}[1]{{\fontsize{9pt}{11pt}\selectfont{\textsf{#1}}}}

\def\A{{\mathbf A}}
\def\B{{\mathbf B}}
\def\C{{\mathbf C}}

\def\I{{\mathbf I}}

\def\M{{\mathbf M}}

\def\S{{\mathbf S}}
\def\t{{\mathbf t}}
\def\u{{\mathbf u}}
\def\v{{\mathbf v}}
\def\W{{\mathbf W}}
\def\X{{\mathbf X}}
\def\Z{{\mathbf Z}}

\def\one{{\mathbf1}}

\def\balpha{\bolds{\alpha}}
\def\bPsi{\bolds{\Psi}}
\def\bpi{\bolds{\pi}}
\def\bSigma{\bolds{\Sigma}}
\def\bepsilon{\bolds{\varepsilon}}
\def\bkappa{\bolds{\kappa}}
\def\bmu{\bolds{\mu}}
\def\bPhi{\bolds{\Phi}}

\def\DN{\mathcal{N}}
\def\IG{\operatorname{IG}}
\def\La{\operatorname{Laplace}}
\def\Exp{\operatorname{Exponential}}
\def\Uni{\operatorname{Uniform}}
\def\Ga{\operatorname{Gamma}}
\def\Wish{\operatorname{Wishart}}
\def\Dir{\operatorname{Dirichlet}}
\def\Dis{\operatorname{Discrete}}

\makeatother

\begin{document}
\begin{frontmatter}

\title{Latent protein trees}
\runtitle{Latent protein trees}

\begin{aug}
\author[A]{\fnms{Ricardo} \snm{Henao}\corref{}\thanksref{t1}\ead[label=e1]{r.henao@duke.edu}},
\author[A]{\fnms{J.~Will} \snm{Thompson}\ead[label=e2]{will.thompson@duke.edu}},
\author[A]{\fnms{M.~Arthur} \snm{Moseley}\ead[label=e3]{arthur.moseley@duke.edu}},
\author[A]{\fnms{Geoffrey S.} \snm{Ginsburg}\ead[label=e4]{geoffrey.ginsburg@duke.edu}},
\author[A]{\fnms{Lawrence} \snm{Carin}\ead[label=e5]{lcarin@ece.duke.edu}}
\and
\author[A]{\fnms{Joseph E.} \snm{Lucas}\thanksref{t2}\ead[label=e6]{joe@stat.duke.edu}}
\runauthor{R. Henao et al.}
\affiliation{Duke University}
\address[A]{Institute for Genome Sciences\\
\quad and Policy (IGSP)\\
Duke University \\
Durham, North Carolina 27708\\
USA \\
\printead{e1}\\
\hphantom{E-mail: }\printead*{e2}\\
\hphantom{E-mail: }\printead*{e3}\\
\hphantom{E-mail: }\printead*{e4}\\
\hphantom{E-mail: }\printead*{e5}\\
\hphantom{E-mail: }\printead*{e6}} 
\end{aug}

\thankstext{t1}{This work was partially done when the author was a
Ph.D. candidate at the Technical University of Denmark.}
\thankstext{t2}{Supported by the Defense Advanced Research Projects
Agency (DARPA), number lN66001-07-C-0092 (G.S.G.).}

\received{\smonth{10} \syear{2012}}
\revised{\smonth{2} \syear{2013}}

%
\begin{abstract}
Unbiased, label-free proteomics is becoming a powerful technique for
measuring protein expression in almost any biological sample. The
output of these measurements after preprocessing is a collection of
features and their associated intensities for each sample. Subsets of
features within the data are from the same peptide, subsets of peptides
are from the same protein, and subsets of proteins are in the same
biological pathways, therefore, there is the potential for very complex
and informative correlational structure inherent in these data. Recent
attempts to utilize this data often focus on the identification of
single features that are associated with a particular phenotype that is
relevant to the experiment. However, to date, there have been no
published approaches that directly model what we know to be multiple
different levels of correlation structure. Here we present a
hierarchical Bayesian model which is specifically designed to model
such correlation structure in unbiased, label-free proteomics. This
model utilizes partial identification information from peptide
sequencing and database lookup as well as the observed correlation in
the data to appropriately compress features into latent proteins and to
estimate their correlation structure. We demonstrate the effectiveness
of the model using artificial/benchmark data and in the context of a
series of proteomics measurements of blood plasma from a collection of
volunteers who were infected with two different strains of viral
influenza.
\end{abstract}

%
\begin{keyword}
\kwd{Proteomics data}
\kwd{hierarchical factor model}
\kwd{coalescent}
\end{keyword}

\end{frontmatter}

\section{Introduction}
Unbiased, label-free, mass spectrometry proteomics,\break sometimes called
``shotgun'' proteomics, is a technique for measuring nearly all
abundant proteins in a biological sample. Because of numerous technical
advances it is becoming increasingly robust and sensitive, leading to
greater effectiveness for the study of biological and medical questions
[\citet{aebersold03,service08,ping09}]. While early work in this field
met with a number of notorious failures [\citet
{petricoin02,baggerly04,zhang05}] due to overlapping peaks, batch
effects and systematic noise, high accuracy spectrometers along with
multiple fractionation techniques such as liquid chromatography and ion
mobility have led to increased robustness as well as improved
qualitative and quantitative results.

After summarization, data generated by this technology is typically
presented as a $p \times n$ dimensional matrix of real-valued
\textit{intensities} where the number of measured features $p$ is typically
orders of magnitude larger than $n$, as in microarray gene expression
data. However, there are a number of important characteristics that
distinguish mass spectrometry proteomics from gene expression data.
First, each feature is a short peptide that has been enzymatically cut
out of a parent protein, and parent proteins typically give rise to
many such peptides. Second, only the more abundant of these features
are typically identified (meaning that the peptide sequence and
originating protein are known). Third, features that are present at
lower abundances will typically have numerous missing values across the
samples. Finally, while the error rate for assigning identifications to
features is low, it is not zero, and this leads to some peptides with
incorrect identifications.

Analysis approaches for these data can be performed at the feature
level or at the protein level. The obvious consequence of performing
analysis at the feature level is a significant loss of power due to the
highly dependent nature of subsets of the features---particularly those
that originate from the same protein. We prefer a dimension reduction
approach in which groups of features are collected and summarized prior
to analysis of associations between features and biological phenotypes.
There are a number of approaches to this in the literature, almost all
of which rely entirely on the identified features in the data set.

The simplest of these approaches involves direct summarization of all
or some features that are identified for each protein either through
averaging or robust summarization based on quantiles [\citet
{poplitya08}]. There are also a number of regression approaches which
include fixed effects for protein, peptide and experimental group
[\citet
{karpievitch09}], include an additional random effect for situations in
which subjects are measured in replicate [\citet{daly08}], or add
additional interaction effects between treatment and features [\citet
{clough09}]. These may assume constant or varying noise levels across
isotope groups and have been shown in some cases to exhibit better
performance than naive summarization approaches that do not adjust for
confounding factors [\citet{clough09}].

We are aware of only one approach to the analysis of these data that
examines correlation structure between data features [\citet{lucas12a}].
This approach utilizes a latent factor model to aggregate features and
uses priors on the loadings that are informed by identifications. This
leads to aggregation of multiple features into ``metaproteins.''
This is a sparse factor modeling approach where nonzero loadings for
factor $i$ are biased toward features that are identified as
originating from protein $i$. While this approach allows the
utilization of unidentified features in the data, it fails to account
for correlation structure that arises when multiple proteins are
involved in the same pathways.

In this paper we present an extension of \citet{lucas12a} that
explicitly models correlation structure between factors. We do this by
incorporating a hierarchical structure on the latent metaproteins that
allows borrowing strength between factors to estimate overall factor
scores. We demonstrate improvements over both a generic sparse factor
model [\citet{carvalho08}] and the earlier proteomics factor model
[\citet{lucas12a}], in terms of accuracy of factor estimates and
eventual association with biological phenotypes. Finally, we
demonstrate the incorporation of known correlation structure in the
form of time series measurements in our analysis of a viral challenge
data set in Section \ref{scflu}.
%
\section{Motivating data} \label{scdata}
While the specifics of data generation may vary at different proteomics
laboratories, the model we describe is appropriate for any
high-accuracy mass spectrometry data. In general, the steps to data
generation are as follows: (i) a biological sample is distilled to a
solution containing those proteins that are of interest; (ii) the
proteins in the sample are then broken up via trypsin; (iii) the
processed sample is separated according to hydrophobicity using liquid
chromatography. The time at which a particular constituent of the
sample passes out of the chromatography column is called the
\textit{retention time}; (iv) an electric charge is induced on the peptides;
(v) the mass and intensity of these ions is measured in a mass
analyzer. The intensity and ion masses are measured at a regular
interval, called the \textit{sampling rate}, and the resulting
measurements form a trace with visible peaks, called \textit{features},
that correspond to one or more peptides. Because the sampling rates are
high relative to the size of these features, each feature spans a range
of mass-to-charge ratios and retention times.

In nature, approximately 1\% of all Carbon atoms are Carbon-13 (they
contain an extra neutron). This leads to multiple features per peptide,
each one containing a different integer number of Carbon-13 atoms.
These are collected into a single \textit{isotope group} (IG) during
preprocessing, and the intensity of this isotope group is estimated as
the total volume under its associated features. In addition to multiple
features from Carbon-13 substitution, a peptide may be present in the
data set multiple times at different charge states. These different
charge states will have different mass to charge ratios and therefore
result in multiple isotope groups per peptide.

There are inherently two different types of correlation present in
label-free, unbiased proteomics data. First, each isotope group
originates from a particular protein and there are typically many
isotope groups per protein in the data set---particularly for proteins
that are highly abundant and/or of large molecular weight in the
original sample. Second, some collections of proteins are expected to
behave similarly because they are part of the same biological pathways.
This will result in correlation between proteins (and therefore
correlation between isotope groups) that are of distinct etiology. In
general, distinct sources of correlation are confounding without some
additional information allowing us to distinguish them. In the case of
proteomics, there are techniques for identifying the specific amino
acid sequence of a subset of the isotope groups that are present at
relatively high concentrations. These sequences are then associated to
particular proteins through sequence alignment to proteins in a
database [\citet{nesvizhskii03}]. We have then, for a limited subset of
the isotope groups, a (possibly imperfect) peptide sequence and
originating protein, which we call an \textit{annotation}.

The proteomics data we will be focused on was obtained from 43 patients
as part of the DARPA H1N1/H3N2 viral challenge project [\citet{zaas09}].
From the entire pool, 24 patients were exposed to H1N1 and 17 were
exposed to H3N2. For each patient, four samples were taken at different
reference time points, baseline ($t=0$), the time of maximum symptoms
($t=1$) as well as $t=0.2$ and $t=0.8$. Each subject was labeled as
symptomatic (SX) or asymptotic (ASX) based on self-reported symptom
scores, as well as viral culture. The samples of the H3N2 study were
run in two batches with the initial pilot study containing only samples
from time points $t=\{0,1\}$ and the followup containing the $t=\{
0.2,0.8\}$ samples. In summary, we have $N=172$ samples from
two studies (H1N1 and H3N2) divided in three batches (H1N1, H3N2$_1$
and H3N2$_2$), two conditions (SX and ASX) where fortunately the
batches and conditions are not confounded. The data itself is a matrix
containing expression values for approximately 40,000 different IGs.
Peptide annotation was done using a combination of Mascot and
PeptideProphet algorithms [\citet{keller02,perkins99a}]. Nearly 85\% of
the IGs remained unannotated. Since H1N1 and H3N2 are two different
experiments, their annotation set is substantially different, thus, an
alignment algorithm must be used in order to take advantage of as much
annotated data as possible, otherwise we will be forced to use only
those IGs shared by both data sets (1697 IGs). Isotope groups from the
three batches were aligned using the algorithm described in \citet
{lucas12a}. From all IGs, 13,845 were successfully aligned across the
H1N1 and H3N2 data sets. From the set of 4670 annotated IGs, only 1697
had annotations in both data sets. The set of annotations consists of
239 proteins from which 106 are assigned to more than one IG. The data
has a relatively low overall missingness rate, most of them among low
abundance IGs. However, missing values are unevenly distributed:
H3N2$_1$ having $10.3\%$ missingness, H3N2$_2$ $0.7\%$ and H1N1 up to
$2.5\%$. Two samples were removed from subsequent analysis because they
had more than $30\%$ missing values in the set of annotated IGs.
%
\section{Model definition} \label{scmd}
We model a sample $n$ of batch $m$ consisting of $p$ IG expressions,
${\mathbf x}
_n^m$, as an extended factor model separated into four effects, namely,
batch, systematic, protein expression and noise,
%
\begin{equation}
\label{eqxAzBwe} {\mathbf x}_n^m = \bmu^m +
\A{\mathbf z}_n + \B{\mathbf w}_n +
\bepsilon_n,
\end{equation}
where ${\mathbf x}_n^m$, $\bmu^m$, ${\mathbf z}_n$, ${\mathbf w}_n$
and $\bepsilon
_n$ are $p\times
1$ vectors. In particular, $\bmu^m$ is the mean expression vector of
batch $m$, factors ${\mathbf z}_n=[z_{1n}\en\cdots\en z_{N_Fn} ]^\top$ are
meant to
capture $N_F$ systematic effects, ${\mathbf w}_n$ is the expression
level of
$N_P$ proteins for sample $n$, $\A$~and $\B$ are $p\times N_F$ and
$p\times N_P$ loading matrices for the systematic effects and protein
expressions, respectively, and $\bepsilon_n$ is measurement
idiosyncratic noise. Systematic effects are included in the model for
the sole purpose of cleaning the data as much as possible from batch
effect specific and technical noise, with the aim to obtain protein
profiles $\{{\mathbf w}_n\}$, that better reflect true biology rather than
technical variability. Provided that protein expression is not directly
observed and because profile vectors $[w_{k1} \en\cdots\en w_{kN}]$ are
likely to be estimated from IGs that belong to multiple proteins, from
now on we refer to them as \textit{latent proteins}. A priori, we let each
IG be associated only to a single latent protein, say, $k$, meaning
that each row of $\B$ contains just one nonzero entry.

Identifiability issues in the model of (\ref{eqxAzBwe}) are
minimized for three reasons: (i)~confounding between systematic effects
and metaproteins is very unlikely because $\A$ is dense and $\B$ is
highly sparse. (ii) ${\mathbf w}_n$ does not have a sign ambiguity because
$\B$
has only nonnegative entries. (iii) ${\mathbf z}_n$ can be identified
up to
scale and permutations as long as its distribution is non-Gaussian [see
\citet{kagan73}]. Scale and permutation ambiguities are not of great
concern here because we are not interested in the interpretation of
systematic effects. Besides, in a case in which batch effects
fully correlate with biological effects, our model will model them
jointly as batch effects. This type of batch confounding is reasonably
common in high-throughput data [\citet{leek10a}], and the failure of our
model to find biological effects when those effects are heavily
confounded with batch is the desired behavior.
%
\subsection{Prior specification}
We need to specify prior distributions for each one of the elements in
the right-hand side of (\ref{eqxAzBwe}). Measurement noise is
set to a zero-mean Gaussian with diagonal covariance matrix $\bPsi$, to
allow for different noise variances for each IG. Entry specific priors
for $\bPsi$ are set to flat inverse gamma distributions with shape
$t_s=1.1$ and rate $t_r=0.001$, the former to keep the variance bounded
away from zero. Mean batch effects have Gaussian priors with mean
$t_m=8$ and small precision $t_p=0.01$, set mainly based on the overall
mean expression of the data. Missing values are provided with
independent standardized Gaussian distributions in order to favor small
values. This reflects the fact that missing values are mostly due to
low abundance peptides.
%
\subsubsection{Systematic effects}
We define systematic effect as a portion of variability expressed in a
large collection of isotope groups that cannot be classified either as
nonspecific measurement noise or biological variability, meaning that
it is more likely due to technical variability. These effects are
usually characterized by high levels of correlation across many isotope
groups, but potentially only in a subset of the samples (e.g.,
only those in one batch). We capture the first part through the use of
independent Gaussian priors on the elements of $\A$, which allows
systematic effects to span the entire set of isotope groups. Aiming to
allow individual samples to be largely dropped from specific systematic
factors, we utilize independent Laplace priors for the elements of
${\mathbf z}
_n$. These are parameterized as scale mixtures of Gaussians with
exponential mixing densities to facilitate inference [\citet{henao11b}].
We consider that the number of systematic factors $N_F$ is not critical
because we are not concerned about the interpretability of matrix $\A$.
Besides, we have observed empirically that the variance explained by
the systematic effect factors saturates quickly as $N_F$ increases.
However, we decided to place an automatic relevance determination (ARD)
prior on $\A$ [\citet{neal96a}]. In particular, being $a_{ij}$ and
$z_{jn}$ elements of $\A$ and ${\mathbf z}_n$, respectively, we have
\begin{eqnarray*}
a_{ij} &\sim& \DN\bigl(0,\rho_j^{-1}\bigr),\qquad
\rho_j \sim \Ga(r_r,r_s),
\\
z_{jn} &\sim& \DN(0,\tau_{jn}),\qquad \tau_{jn}
\sim \Exp\bigl(\lambda^2\bigr),\qquad \lambda^2 \sim
\Ga(\ell_s,\ell_r),
\end{eqnarray*}
where $\rho_j$ is a shared factor-wise variance for the columns of $\A$
and $\tau_{jn}$ is an auxiliary variance with exponential mixing so,
marginally, $z_{jn} \sim\La(\lambda^2)$ [\citet{andrews74}]. We further
place a gamma hyperprior on the rate of the Laplace distribution with
parameters $\ell_s=4$ and $\ell_r=2$. The ARD is a variable selection
prior; Large values of $\rho_j$ will correspond to small values of the
$j$th column of $\A$, thus virtually \textit{switching off} the entire
effect. Setting $r_r=1.1$ and $r_s=0.001$ will encourage the desired
behavior. In practice, the \textit{effective} number of factors can be
determined by thresholding $\rho_j$ or the elements of $\A$ column-wise.
%
\subsubsection{Latent protein profiles}
We make two assumptions regarding isotope group expression. One is that
each isotope group originates from only one latent protein and the
other is that latent proteins may correlate with each other due to
biological pathway activity. To model the first feature, we set a prior
hierarchy as follows:
\[
b_{i,u_i}|u_i \sim\DN_+(0,1),\qquad u_i|
\v_i \sim\Dis(\v_i),\qquad \v_i|\alpha
\sim\Dir(\alpha\one_{N_P}),
\]
where $b_{i,j}=0$ if $j\neq u_i$, $\DN_+(\cdot)$ is the Gaussian
distribution truncated below zero and where the $i$th IG is associated
with the latent protein indexed by $u_i$ with probability $\v_i$. This
means that vector $\u$ serves as a labeling variable for IGs. The
conjugate prior for the vector of $N_P$ probabilities, $\v_i$, is set
using a shared concentration~$\alpha$. For the latter, we provide a
flat gamma prior with parameters $a_s=1$ and $a_r=1$ [see
\citet{escobar95a}].

\begin{figure}

\includegraphics{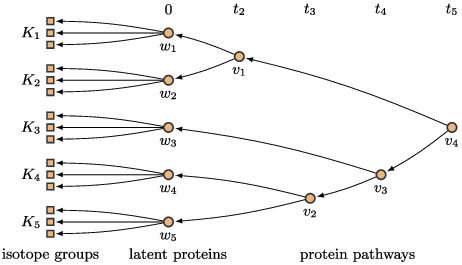}

\caption{Latent protein tree structure.
Particular tree with $N_P=5$ and three isotope groups assigned to each
latent protein. The pseudo time variable $t$ defines the merging points.}
\label{fgptree}
\end{figure}

We know that groups of proteins might have similar expression profiles
for different reasons, for example, because they are structurally
similar, mediate similar biological processes, share a pathway, etc. In
order to capture this structure, we place a prior over binary trees on
the $N_P$ latent proteins. This allows us to model correlation among
metaproteins and leads to an interpretable representation of isotope
groups, latent proteins and their interactions. Figure \ref{fgptree}
illustrates the concept for a particular setting with $p=15$ IGs
distributed in $N_P=5$ proteins. We can see a hierarchical clustering
structure in which, for instance, latent proteins $w_1$ and $w_2$ are
more similar than $w_4$ and $w_5$, thus more correlated. The
\textit{pseudo time} $t_j$ at which two nodes merge into $v_j$ acts as a
similarity measure so that more alike latent proteins merge sooner in
time, allowing us to directly quantify their pairwise or group-wise
similarities. The proposed hierarchy is an implementation of the
Kingman's coalescent [\citet{kingman82}] and reflects the idea that
isotope groups and latent proteins lay in different levels and that
protein pathways are proxies for the average profiles of collections of
proteins.

Given a tree structure, $\{\t,\bpi\}$, where $\t$ is the vector of
merging times and $\bpi$ is the set of partitions at each level of the
tree, we specify the relationship between node $v_j$ and its parent
node $n_k$ (or $w_k$ at the leaves) through a multivariate Gaussian
transition probability and set the following prior hierarchy:
%
\begin{equation}
\label{eqtree}\quad \v_j|\v_k,t_j,t_k,
\bPhi\sim\DN\bigl(\v_k,(t_k-t_j)\bPhi
\bigr),\qquad \{\t,\bpi\} \sim\operatorname{Coalescent}(N_p),
\end{equation}
where $\v_j$ is a $N$-dimensional row vector and $\bPhi$ is a
covariance matrix encoding the correlation structure in $\v_j$. A
coalescent prior selects a pair to merge uniformly from partition $\pi
_j$ and sets merging times with rate 1, this is $t_k \sim\Exp(1)$. With
no further constraints, this prior distribution leads to a uniform
prior distribution over trees that is independent of merging times and
is infinitely exchangeable [Kingman (\citeyear{kingman82,kingman82a})].
Different priors for $\bPhi$ add flexibility to the model, for example,
in the i.i.d. case, a~diagonal $\bPhi$ with independent inverse gamma
prior distributions on each diagonal element will accommodate for
differing levels of noise for different samples. In cases where there
is known structure, a different prior could be used. In our analyses we
use inverse Wishart priors to model correlation due to sample
replicates and Gaussian process priors for smoothness in time series
data. Inference for hierarchy in (\ref{eqtree}) is carried out using an
efficient sequential Monte Carlo Sampler introduced by
\citet{henao12b}.

\subsection{Inference}
Model fitting is performed using Markov chain Monte Carlo (MCMC) to
collect samples from the posterior of all parameters in the model,
namely, $\bmu^m$, $\A$, ${\mathbf z}_n$, $\B$, ${\mathbf w}_n$,
$\bPsi$,
$\u$, $\bpi$ and
$\bPhi$. The most relevant summaries involve posterior samples from the
latent proteins, IG-protein assignments and the hierarchical structure
encoded by the binary tree, $\bpi$. Nearly all quantities of interest
are updated using Gibbs sampling except for the tree components that
require sequential Monte Carlo (SMC) sampling. In all the experiments
described in this paper we set the hyperparameters of the model to the
values already mentioned unless otherwise stated. The upper bound for
the number of factors is set to a conservatively large value; we have
observed in practice that $N_F=\lfloor2\log(p)\rfloor$ is large enough.
For tasks with $p$ and $N$ in the lower thousands and hundreds,
respectively, we can expect the inference routine to take less than a
couple of hours in a desktop machine. The entire sampling sequence is
fully described in the \hyperref[app]{Appendix}.

Summaries for most of the important quantities of the model are
computed in the usual way by means of histograms and empirical
quantiles. Summarizing trees, on the other hand, is not such an easy
task because tree averaging is not a well-defined operation. We could,
in principle, use the pseudo time variable to build a pairwise distance
matrix between latent proteins and then attempt to build a tree from a
summary of such a \textit{similarity} matrix. The problem being that we
do not have any guarantee that this \textit{average} of binary trees will
produce a binary tree as well. We tried this approach with both
artificial and real data, and found that the tree built using means or
medians of the similarity matrices collected during inference
oftentimes produced trees with nonbinary branching, thus not matching
the prior assumption. In view of this, we decided to select a single
tree from all the available samples using as criterion the marginal
likelihood of the tree. This is a common practice in tree based models;
see, for instance, \citet{teh08} and \citet{adams10}.

The source code and demo scripts for the model presented in this paper
are written in MATLAB and C, and have been made publicly available at
\url{http://www.duke.edu/\textasciitilde rh137/files/lpt_v0.3.tar.gz}.
%
\section{Artificial data}
We begin with a set of experiments using artificially generated data in
order to illustrate some of the features of our model and to perform
some quantitative comparisons. We generated two data sets $D_1$
and $D_2$ of sizes $\{p,N,N_B,N_F,N_P\}=\{800,80,2,4,32\}$ and $\{
1600,80,3,6,64\}$, respectively. Denoting the elements of $\bmu^m$,
$\A
$, $\B$ and $\bPsi$ as $\mu_i^m$, $a_{ij}$, $b_{ik}$ and $\psi_i$,
respectively, we draw $N$ observations of the model from the following
hierarchy:
\begin{eqnarray*}
{\mathbf x}_n^m &\sim& \DN\bigl(\bmu^m,
\bSigma\bigr),
\\
\mu_i^m &\sim& \DN(8,2),\qquad m \sim\Dis
\bigl(N_B^{-1}\one_{N_B}\bigr),
\\
a_{ij} &\sim& \DN(0,0.1),
\\
b_{i,u_i} &\sim& \DN_+(0,1),\qquad u_i \sim\Dis(\v),
\\
\psi_i^{-1} &\sim& \Ga(1.1,0.02),\qquad \v\sim\Dir(\balpha),
\\
\S^{-1} &\sim& \Wish(\I,N_P),\qquad \balpha\sim\Uni(0.8,2.4),
\end{eqnarray*}
where $\bSigma=\A\A^\top+ \B\S\B^\top+ \bPsi$, $\A$ is a
$p\times N_F$
matrix of systematic factor loadings, $\B$ is a $p\times N_P$ matrix of
latent protein loadings, $\S$ is the covariance matrix of the latent
protein profiles and $\bPsi$ is the noise diagonal covariance matrix,
as in (\ref{eqxAzBwe}). We generated 50 replicates of
each data set and uniformly flagged 20\% of its values as missing. We
ran our sampler for 4000 iterations, using the first 3000 as burn-in
period. For this experiment, we set the distribution of the systematic
factors to Gaussian, to match the assumption made in $\bSigma$. Since
we are not introducing correlation across samples, we set $\bPhi$ to
diagonal with independent gamma priors. The average number of
systematic factors is selected with threshold $\rho_j<10^{3}$. We label
each latent protein by tabulating the IGs associated to it from vector
$\u$ and then picking the label having maximum count. We define
\textit{identity} as the percent of correctly labeled latent proteins and
\textit{confusion} as the percent of variables incorrectly associated to their
latent proteins. We compare our model (LPT) with (i) its simplified
version without the tree structure inference we call sLPT, thus without
covariance structure in the latent profiles [\citet{lucas12a}].
Table \ref{tbartificialnf} shows results for the structural
components of the model---identity, confusion and number of systematic
factors. Results demonstrate that the model is able to capture the
association between IGs and latent protein profiles through $\u$ while
properly handling ``batch'' effects and missingness in the data.
The two methods perform similarly because estimates of systematic
effects and peptide-protein associations is only weakly influenced by
the protein tree structure. Even so, LPT performs slightly better than
sLPT in terms of protein association accuracy.

%
\begin{table}
\tablewidth=300pt
\caption{Structural measures for artificial data. $N_F$ is selected
with threshold $\rho_j<10^{3}$. Pairs in brackets are empirical 90\%
intervals across replicates. Best results in boldface letters}
\label{tbartificialnf}
\begin{tabular*}{\tablewidth}{@{\extracolsep{4in minus 4in}}lclll@{}}
\hline
\textbf{Set} & \textbf{Method} & \multicolumn{1}{c}{$\bolds{N_F}$}
& \multicolumn{1}{c}{\textbf{Identity}} & \multicolumn{1}{c@{}}{\textbf{Confusion}} \\
\hline
$D_1$ & LPT & \textbf{4} $\bolds{(3,7)}$
& \textbf{0.97} $\bolds{(0.94,1.00)}$ &
\textbf{0.002} $\bolds{(0.000,0.009)}$ \\
& sLPT & \textbf{4} $\bolds{(3,7)}$ & 0.97 $(0.91,1.00)$ & 0.005 $(0.000,0.016)$ \\
$D_2$ & LPT & \textbf{6} $\bolds{(5,10)}$ & \textbf{0.98} $\bolds{(0.97,1.00)}$
& \textbf{0.003} $\bolds{(0.00,0.008)}$ \\
& sLPT & 6 $(5,10)$ & 0.97 $(0.93,1.00)$ & 0.007 $(0.001,0.014)$ \\
\hline
\end{tabular*}
\end{table}

We can also assess the performance of our model in terms of covariance
matrix and missing value estimation. We compare LPT and sLPT as well as
a sparse factor model as proposed by \citet{carvalho08}, sFM,
which utilizes the same priors for missing values and batch effects
used by our model. For sFM we set the number of factors to
$N_F+N_P=\{21,24\}$, accordingly. In principle, the sparse model is
flexible enough to estimate $\A$ and $\B$ but not $\S$, for the model
assumes independent profiles, similar to sLPT. Table~\ref{tbartificialmse} shows summaries of mean square error
(\textsc{mse}), mean absolute error (\textsc{mae}) and maximum absolute
bias (\textsc{mab}) across replicates for the methods under
consideration. As seen in Table \ref{tbartificialmse}, our model
performs better than the other two alternatives. In particular, we see
that sLPT and LPT behave similarly in terms of missing value
estimation, however, LPT significantly outperforms the others in terms
of covariance matrix estimation, as the model explicitly accounts for
it. Significance is measured in terms of median \textsc{mse},
\textsc{mae} and \textsc{mab} pairwise differences with $p$-value
threshold $0.01$.

%
\begin{table}
\caption{Performance measures for artificial data. sLPT is the
simplified LPT and sFM is a sparse factor model. \textsc{mse}, \textsc{mae}
and $10^{-1}\times$\textsc{mab} are mean squared error, mean absolute
error and maximum absolute bias, respectively. Pairs in brackets are
empirical 90\% intervals. Best results shown in boldface letters.
Differences in covariance measures between LPT and sLP are
significant with $p$-value threshold $0.01$} \label{tbartificialmse}
\begin{tabular*}{\tablewidth}{@{\extracolsep{\fill}}lclll@{}}
\hline
\textbf{Set} & \multicolumn{1}{c}{\textbf{Measure}} & \multicolumn{1}{c}{\textbf{LPT}}
& \multicolumn{1}{c}{\textbf{sLPT}} & \multicolumn{1}{c@{}}{\textbf{sFM}} \\
\hline
\multicolumn{2}{@{}l}{Covariance} & \\
\quad$D_1$ & \textsc{mse} & \textbf{1.291} $\bolds{(0.898,1.678)}$ & 4.538
$(2.813,7.738)$ & 4.776 $(3.029,7.673)$ \\
& \textsc{mae} & \textbf{0.883} $\bolds{(0.748,1.016)}$ & 1.472 $(1.217,1.922)$ & 1.396
$(1.179,1.874)$ \\
& \textsc{mab} & \textbf{0.753} $\bolds{(0.532,2.287)}$ & 1.204 $(0.939,2.454)$ & 1.473
$(1.176,7.703)$ \\
\quad$D_2$ & \textsc{mse} & \textbf{1.143} $\bolds{(0.978,1.525)}$ & 2.439
$(1.922,3.381)$ & 2.434 $(2.018,3.683)$ \\
& \textsc{mae} & \textbf{0.840} $\bolds{(0.787,0.946)}$ & 1.079 $(0.974,1.286)$ & 1.001
$(0.865,1.182)$ \\
& \textsc{mab} & \textbf{0.848} $\bolds{(0.636,4.844)}$ & 1.161 $(0.996,4.958)$ & 1.658
$(1.163,8.871)$ \\
\multicolumn{2}{@{}l}{Missing values} & \\
\quad$D_1$ & \textsc{mse} & \textbf{0.144} $\bolds{(0.083,0.352)}$ & 0.150
$(0.088,0.376)$ & 1.935 $(1.221,2.514)$ \\
& \textsc{mae} & \textbf{0.193} $\bolds{(0.178,0.215)}$ & 0.195 $(0.179,0.212)$ & 0.690
$(0.536,0.845)$ \\
& \textsc{mab} & \textbf{0.850} $\bolds{(0.473,2.908)}$ & 0.890 $(0.586,2.902)$ & 1.096
$(0.939,2.347)$ \\
\quad$D_2$ & \textsc{mse} & \textbf{0.146} $\bolds{(0.110,0.367)}$ & 0.148
$(0.105,0.341)$ & 2.345 $(1.894,2.933)$ \\
& \textsc{mae} & \textbf{0.193} $\bolds{(0.184,0.211)}$ & 0.194 $(0.184,0.213)$ & 0.784
$(0.679,0.913)$ \\
& \textsc{mab} & 1.102 $(0.724,2.936)$ & \textbf{1.018} $\bolds{(0.727,2.426)}$ & 1.200
$(1.040,2.537)$ \\
\hline
\end{tabular*}
\end{table}

The entire experiment was repeated for small variations in the
hyperparameters of the models and the artificial data generator without
considerable changes in the results. In general terms, we observed good
mixing in the sampler using exploratory and standard diagnostic tests.
We also repeated the experiment with correlation across samples and an
inverse Wishart distribution for the matrix $\bPhi$ with results
similar to those in Tables \ref{tbartificialnf} and \ref{tbartificialmse}.
%
\section{Confounding due to batches}
\begin{figure}

\includegraphics{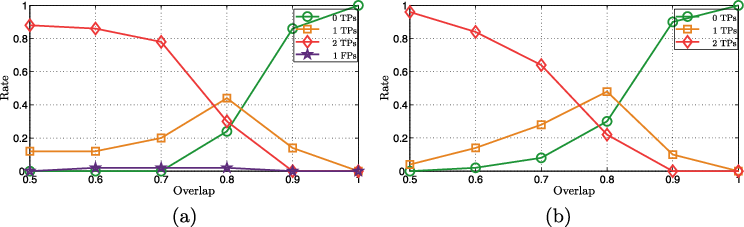}

\caption{Confounding effects results for $D_1$
\textup{(a)} and $D_2$ \textup{(b)}. Each marker represents the proportion of replicates
(50) for which our model found 0, 1, 2 (ground truth) positives and
false positives. Mind that rates for true positives sum up to 1.}\label
{fgconfoundd1}
\end{figure}

Next we explore how different levels of confounding between
biological and batch effects impact results. For this purpose, we
generated 50 replicates of a modified version of data sets $D_1$ and
$D_2$ from a previous experiment in which we set $N_B=2$ and added 2
\textit{biological effects} as follows:
\[
w_{1n},w_{2n} \sim\DN(\mu_e,1),\qquad
w_{kn} \sim\DN(0,1),
\]
where $\mu_e=0.75$ or $\mu_e=-0.75$ if sample $n$ has a
\textit{positive} or \textit{negative} biological effect, respectively, and
$k=3,\ldots,\{32,64\}$. Batch indicators are drawn uniformly, but
biological effect indicators are obtained such that a proportion ($\tau
$) of samples share both indicators. When $\tau=0.5$ the overlap is
minimum and when $\tau=1$ batch and biological effects are fully
confounded, as both can be jointly captured as batch means. For the
results, we computed the proportion of times our model found 0, 1, 2
(ground truth) true positives and 1, 2, etc. false positives.
Biological effects are tested for on each protein using $t$-tests with
$p$-value threshold $0.01$ and Bonferroni correction for the number of
proteins. Figure \ref{fgconfoundd1}(a) shows that for the minimum
%
%
\begin{figure}[b]

\includegraphics{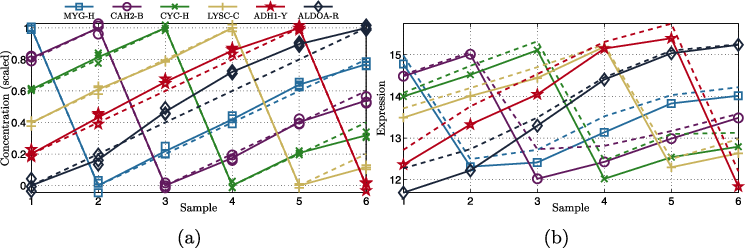}

\caption{Spike-in data profiles. \textup{(a)} Ground
truth (dashed) and estimated (solid) protein profiles scaled between 0
and 1. Replicates are shown as markers and solid lines are averages
across replicates. \textup{(b)} Median IG expression grouped according to the
labeling obtained during inference and averaged across replicates.
Dashed lines correspond to original data with missing values and solid
lines to data with missing values replaced by their estimates. Credible
intervals were omitted for clarity.}\label{fgspikeinrefest}
\end{figure}
overlap our model finds the 2 biological effects approximately 90\% of
the times and that such a proportion decreases to exactly zero (100\% 0
true positives) as the $\tau$ approaches 1. We also see that the false
positive rate is very small and that for large overlaps is always zero.
As the model is currently defined, any effect that correlates with
batch indicators will be treated as a batch effect, in that sense,
confounded biological effects cannot---and arguably should not---be
detected.

\section{Spike-in data}
The benchmark data set originally introduced by \citet{mueller07a}
consists of 6 samples measured in three replicates. Each sample is a
mixture of six nonhuman purified proteins in different concentration
levels spanning two orders of magnitude from 25 to 800 fmol.
Figure \ref{fgspikeinrefest}(a) shows, in dashed lines, ground truth
concentrations on a log-scale and scaled to fit in the interval $[0,1]$.
The raw data containing approximately 15,900 IGs per sample was
filtered down to 1841 IGs per sample after identification, annotation
and exclusion of unidentified IGs with 50\% missing values or with less
than 10\% of the maximum variance IG. Annotations are available for
only 88 IGs; This is 4.7\% of the set. The final data set contains 18
observations and 1841 IGs labeled with 7 protein names, \lp{ADH1-Y}
(12), \lp{ALDOA-R} (20), \lp{CAH2-B} (13), \lp{CYC-H} (24), \lp{LYSC-C}
(9), \lp{MYG-H} (10) and \lp{UKN} (1753), with the number of IGs per
protein in parentheses and \lp{UKN} denoting unannotated IGs. The data
matrix has a missingness of 30\% that is more or less evenly
distributed across observations. The original experiment reported by
\citet{mueller07a} only uses annotated data. Since the data set is
relatively clean and all the samples were obtained in a single session,
we do not expect systematic, batch effects or a meaningful covariance
structure. However, we do expect high correlation due to replicates,
thus, we provide $\bPhi$ with an inverse Wishart prior with $10\times
N$ degrees of freedom and scale matrix composed of 6 blocks of
magnitude 0.9 and size 3 plus 0.1 times the identity matrix. Although
learning the degrees of freedom and the blocks/diagonal proportions
will be more principled, we did not observe substantial changes in the
results from small changes in the previously mentioned values. We ran
the sampler for 4000 iterations with a burin-in period of 2000.

Figure \ref{fgspikeinrefest}(a) shows the summary of the estimated
latent protein profiles. Each circle represents a replicate, solid
lines are averages across replicates and dashed lines represent the
ground truth [see \citet{mueller07a}]. Summaries were computed using
medians and credible intervals were omitted for clarity.
Summaries with credible intervals are available as
the supplementary material [\citet{Henetal13N3}].
Compared to the ground truth, our model does a pretty good job at
capturing the underlying profiles of all 6 proteins of interest despite
the large amount of missing values and unannotated IGs used.

Availability of the true protein profiles allows us to
quantitatively evaluate how accurate our model is at estimating the
protein profiles. We compare four different models: (i) the model for
protein quantitation described in \citet{karpievitch09} where we have
used protein concentrations as a grouping variable (Karp09) and three
variants of our model, (ii) full i.i.d. latent proteins, meaning no tree
structure prior; (iii) independent gamma distributions and diagonal
$\bPhi$, assumes no correlation due to replicates and (iv) inverse
Wishart prior for $\bPhi$ with scale matrix as already described.
Results of model (iii) also appear in \citet{henao12a}. Although the
three factor models [(ii)--(iv)] produce profiles similar to those shown in
Figure~\ref{fgspikeinrefest}(a), there are small differences.
Table \ref
{tbspikeinmse} indicates that in terms of \textsc{mse}, \textsc{mae} and
\textsc{mab}, the results of the model with the inverse Wishart prior (iv)
are most accurate. Although the covariance structure in the
true protein profiles is not interpretable in this experiment, they are
correlated, which explains why the two models with tree structure prior
[(iii) and (iv)] outperform the full i.i.d. models [(i)~and (ii)]. Additionally,
the inverse Wishart prior in model (iv) is improved over model (iii)
because the prior accounts for the sample correlation resulting from
having replicates in the experiment.

\begin{table}
\tablewidth=270pt
\caption{Performance measures for spike-in data. \textsc{mse}, \textsc{mae}
and \textsc{mab} are mean squared error, mean absolute error and maximum
absolute bias, respectively} \label{tbspikeinmse}
\begin{tabular*}{\tablewidth}{@{\extracolsep{\fill}}ld{2.3}ccc@{}}
\hline
& & & \multicolumn{2}{c@{}}{\textbf{Tree with} $\bPhi$
\textbf{prior}} \\[-4pt]
& & & \multicolumn{2}{c@{}}{\hrulefill}\\
\textbf{Measure} & \multicolumn{1}{c}{\textbf{Karp09}} & \textbf{No tree}
& \textbf{Indep. gamma} & \textbf{Inverse Wishart} \\
\hline
$10^3\times\mbox{\textsc{mse}}$ & 12.370 & 2.524 & 1.899 & 1.661 \\
$10^2\times\mbox{\textsc{mae}}$ & 6.915 & 3.172 & 2.983 & 2.494 \\
$10^1\times\mbox{\textsc{mab}}$ & 3.094 & 1.443 & 1.252 & 1.213 \\
\hline
\end{tabular*}
\end{table}

We can use the labeling vector $\u$ to examine how unannotated isotope
groups were labeled after inference. In particular, \lp{ADH1-Y} went
from having 12 IGs to 118, \lp{ALDOA-R} from 20 to 307, \lp{CAH2-B}
from 13 to 240, \lp{CYC-H} from 24 to 288, \lp{LYSC-C} from 9 to 189
and \lp{MYG} from 10 to 185. Figure \ref{fgspikeinrefest}(b) shows
median IG expression grouped according to the labeling vector $\u$ and
averaged across replicates to make easier comparisons against the
ground truth in Figure \ref{fgspikeinrefest}(a). Dashed and solid lines
correspond to data with and without missing values, respectively.
For the latter, we have replaced the missing values with those
estimated by our model. We see that for every protein our model
estimates of missing values improve the expression average. The largest
improvement is in the lower end of the expression range, precisely
where the missing values are likely to be found [see \citet
{mueller07a}]. A similar picture using only the labeling from
annotation does not resemble the ground truth at all. This is because
the original labeling only comprises 88 IGs with a considerable amount
of missing values.
%
%
\section{H1N1/H3N2 viral challenge} \label{scflu}
We present now the case study based on the motivating data already
described in Section \ref{scdata}. Here we will be using only the set
of 4670 annotated IGs for which we have at least 2 IG per protein.
Therefore, for this study we have $n=172$, $N_B=3$, $N_F=16$ and
$N_P=106$. Additionally, each observation can be seen as an element of
a time series of length 4, that is, $t=\{0,0.2,0.8,1\}$. If we let
latent proteins have Gaussian process priors with squared exponential
covariance function and assuming no sample correlation across patients,
we can compute the entries of $\bPhi$ from
\[
\phi(i,j) = c_{ij}\exp\bigl(-
\ell^{-1}d_{ij}^2 \bigr) + \sigma^2
\delta_{ij},
\]
where $\ell$ is the inverse length scale, $\sigma^2$ the idiosyncratic
noise variance, $\delta_{ij}=1$ only if $i=j$, $c_{ij}=1$ only if
samples $i$ and $j$ are from the same patient, and $d_{ij}=t_i-t_j$ is
the time difference between pair $\{t_i,t_j\}\in\{0,0.2,0.8,1\}$.
Hyperparameters $\ell$ and $\sigma^2$ are updated using slice sampling
[\citet{neal03a}]. We ran the inference procedure for 5000 burn-in
iterations followed by 2000 samples to compute summaries. The whole
procedure takes approximately 2.5 hours in a regular desktop machine
with 4 cores. Mixing was monitored using both exploratory and standard
diagnostic tests. Table \ref{tbhxnxnf} reports the resulting
%
\begin{table}
\tablewidth=205pt
\caption{Structural measures for viral challenge data. $N_F$ is
selected with threshold $\rho_j<10^{3}$ and stability with threshold
$0.6$} \label{tbhxnxnf}
\begin{tabular*}{\tablewidth}{@{\extracolsep{\fill}}lcccc@{}}
\hline
$N_F$ & Identity & Confusion & Stability & Unique \\
3 & 0.774 & 0.511 & 0.958 & 0.783 \\
\hline
\end{tabular*}
\end{table}
structural components of the model, namely, previously described:
number of systematic factors, $N_F$, identity and confusion. We define
\textit{stability} as the proportion of IGs having a single value in the
label vector $\u$ for at least 60\% of the MCMC samples after the
burn-in period. We also define \textit{unique} as the proportion of latent
proteins with distinct labels.

\subsection{Consistency with annotation}
From Table \ref{tbhxnxnf} we see that approximately half of the IGs
ended up with a protein label different from their annotation
(\textit{confusion}). Possible explanations for this include systematic
effects, post-translational modifications, measurement error and
alignment induced mislabeling. In this example, consider the problem of
aligning batches H1N1, N3N2$_1$ and N3N2$_2$. Initially, the three
batches have different sets of annotation that need to be matched to
create a common annotation set. We use the alignment algorithm
described in \citet{lucas12a}. From the 4670 IGs included in the model,
annotation was transferred from one of the batches to the other two in
$64\%$ of the cases. This means that more than half of the IGs are more
prone to miss-annotation due to the challenges of aligning between data
sets. We found that a disproportionate percentage of peptides that
retained their label from annotation after model fit are from the set
of IGs with H1N1/H3N2 shared annotation. This suggests that IGs
annotated simultaneously in all sets tend to be more reliable than
those labeled by label transfer.

\begin{figure}

\includegraphics{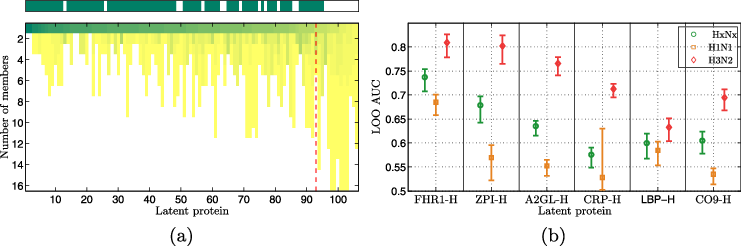}

\caption{Protein identification and status
classification. After model fitting, each latent protein contains a set
of peptides, not all of which are from the same protein. \textup{(a)} Number of
members or protein labels per latent protein. Each column is a
different latent protein. For a particular column, each row contains
membership information, ordered top to bottom from most to least common
for the corresponding latent protein. Color encodes member dominance,
thus, dark green indicates that a given latent protein is dominated by
peptides annotated by protein prophet as originating from a single
protein. The red line separates latent proteins in which the leading
member has a proportion less than 30\%. The top bar shows in dark the
82 proteins whose posterior label matches prior information. \textup{(b)}
Classification accuracy presented as AUC values estimated using
leave-one-out cross-validation. Markers indicate median values and
error bars cover $90\%$ credible intervals.}\label{fgpurity}
\end{figure}

The identity of the model, on the other hand, indicates that 82 latent
proteins match annotation when labeled by consensus of their IG
members. The remaining latent proteins represent cases of duplicate
representation of particular proteins. For example, there are 6
latent proteins associated with \lp{APOB-H} (the most commonly
identified protein in the data), all of them with disparate profiles.
Figure~\ref{fgpurity}(a) shows the composition of all latent proteins.
%
For each latent protein (column), we tabulate and sort the
labels of its IG members (rows). Darker colors represent proportions
closer to 1. The first row is used to compute the consensus to
determine identity. The red bar indicates whether the most frequent IG
in a given latent protein is represented by less than 30\% of the IGs
assigned to it. The top bar shows in dark the 82 latent proteins that
match their initial annotation. For most latent proteins, the most
frequent IG has an important contribution and no latent protein has IGs
from more than 17 different labels.

\subsection{Association with phenotype and pathway analysis}
We can also use latent proteins as predictors of the symptomatic vs.
asymptomatic status of each observation in the data set. For this
purpose, we fit individual linear discriminant classifiers for each
latent protein at each MCMC draw and estimate the classification
accuracy as the area under the ROC curve [AUC, Receiver Operating
Characteristic, \citet{fawcett06}]. Figure \ref{fgpurity}(b) shows results
for the six most discriminant latent proteins: \lp{FHR1-H}, \lp{ZPI-H},
\lp{CRP-H}, \lp{LBP-H}, \lp{A2GL-H} and \lp{CO9-H}; It shows in
particular that \lp{FHR1-H} has an overall decent performance. In
addition, when treating H1N1 and H3N2 as separate classification tasks,
we observe that H3N2 is clearly easier to classify.\looseness=1

We also applied the model for protein quantitation of \citet
{karpievitch09} using symptomatic/asymptomatic status as a grouping
variable. Their model found 40 significant proteins with $q$-value
threshold 0.05, which is quite a large number considering the total
number of proteins in the data set is 106. In addition, almost none of
these show significant association with the biological phenotype. We
found only 3 proteins in common (\lp{CHLE-H}, \lp{FHR1-H} and \lp
{HRG-H}) when comparing their list to our own. For our model we used
$t$-tests, $q$-values and the same 0.05 threshold to be fair with the
other method. However, their list does not include \lp{ZPI-H}, \lp
{CRP-H}, \lp{LBP-H}, \lp{A2GL-H} or \lp{CO9-H}, all of which are
strongly associated with the symptomatic versus asymptotic designation.

As described in Section \ref{scmd}, the prior distribution for the set
of latent proteins allows us to build a binary tree representation of
its elements in a hierarchical clustering fashion. When examining the
resulting structure [see \citet{Henetal13N1}] we found some straightforward
groupings in the tree mostly corresponding to protein variants like \lp
{APOC2-H} and \lp{APOC3-H}, \lp{CO8A-H}, \lp{CO8B-H} and \lp{CO8G-H},
\lp{FIBG-H} and \lp{FIBB-H}, \lp{F13A-H} and \lp{F13B-H}, etc., all of
them having similar profiles when looking at their estimated signatures
(results not shown), in other cases, for instance, \mbox{\lp{CO4(a,b)-H}} and
\lp{APOB-H}, showing great diversity in their profiles and as a result
rather spread in the structure.

\begin{figure}

\includegraphics{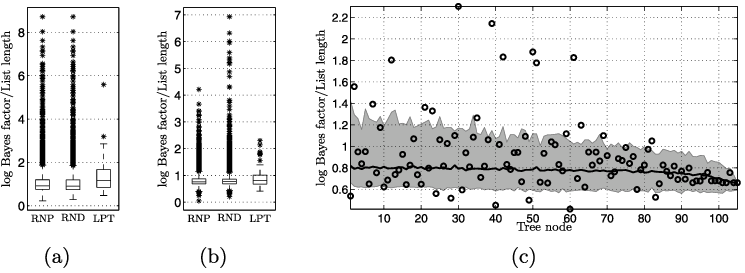}

\caption{GO scaled log Bayes factors. \textup{(a)} Latent
proteins. \textup{(b)} Tree nodes. \textup{(c)} Bayes factors vs tree nodes for LPT
(circles) and RNP (solid line). Shaded area covers 90\% empirical
quantiles for RNP values.}\label{fglpbf}
\end{figure}

In an attempt to quantify whether the latent proteins and tree
representation produced by our model is meaningful from a biological
point of view, we performed Gene Ontology (GO) searches for the protein
lists encoded by each latent protein and each tree node. In order to
quantify the strength of the association between GO annotations and our
protein lists, we use Bayes factors [GATHER, \citet{chang06a}]. As
controls we generated (i)~500 latent proteins/trees from the prior in
(\ref{eqtree}) (RND) and (ii) 500 random label permutations
for the latent proteins and tree produced by our model (RNP). Figure
\ref{fglpbf}(a) and (b) shows separate Bayes factor
%
boxplots for
latent proteins and tree nodes, respectively. Bayes factors have been
scaled by the size of the protein list to compensate for the
agglomerative mechanism of the tree structure. Differences in medians
between LPT and the two controls are significant with $p$-value
threshold 0.01 for both latent proteins and tree nodes. Provided that
LPT and RNP have the same tree structure, we can directly compare Bayes
factors at each node of the tree. Figure \ref{fglpbf}(c) shows scaled
Bayes factors for each tree node of LPT (circles) and RNP (median:
solid line; shade: 90\% empirical quantiles). We see quite a few nodes
with Bayes factors far exceeding the domain of randomly permuted
protein labels. These nodes are the ones with a high level of evidence
for association with the GO annotations complement activation, immune
response, acute-phase response, cytolysis and response to pathogen. The
node with largest Bayes factor [node 30 in Figure \ref{fglpbf}(c)]
contains \lp{CRP-H} and \lp{LBP-H}, two of our most predictive latent
proteins.

\begin{figure}

\includegraphics{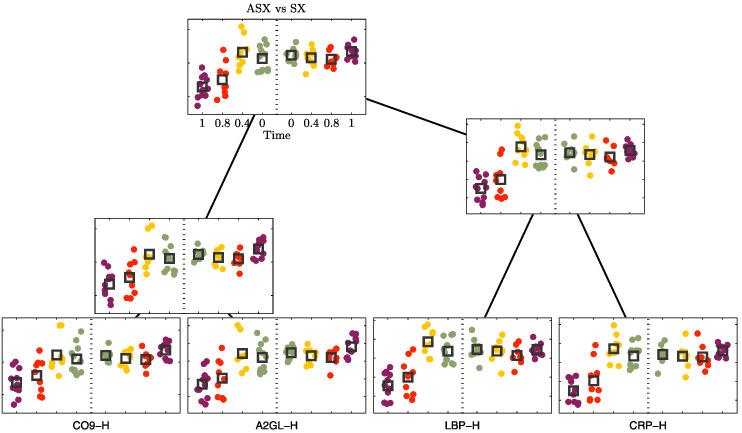}

\caption{Discriminant subtree. This figure
shows a set of three internal nodes and four leaves from the latent
protein tree structure. Each node is represented as a scatter plot
showing samples (dots) from the H3N2 study. The vertical dotted line
separates asymptomatic (left) and symptomatic (right) samples. Samples
are grouped along the $x$-axis according to time stamp: green for $t=0$
(closest to dashed line), yellow for $t=0.2$, red for $t=0.8$ and
purple for $t=1$ (farthest toward the outside edge). The $y$-axis is
the estimated latent/protein pathway expression. The mean for each
group and time point is denoted with a square. For this group of latent
proteins, the symptomatic subjects at time points $t=0.8$ and $t=1$
show clear separation.}
\label{fgstree}
\end{figure}

Figure \ref{fgstree} shows the subtree corresponding to 4 of the
discriminant proteins from Figure \ref{fgpurity}(b) along with a
scatter of
the expression values of each latent protein. Each panel in the figure
shows expression in the $y$-axis and data grouping in the $x$-axis.
Data to the left-hand side of the dashed vertical line corresponds to
the asymptomatic set, whereas the other side contains symptomatic
observations. Each side is further grouped according to time, so points
closer to the dashed vertical line are for $t=0$ (green), then $t=0.2$
(yellow), $t=0.8$ (red) and the farthest to the outside is $t=1$
(purple). The good separation of observations from times $t=\{0.8,1\}$
is the feature responsible for the classification results shown in
Figure~\ref{fgpurity}(b). The node above \lp{CRP-H} and \lp{LBP-H}
in Figure \ref{fgstree} is node 30 in Figure \ref{fglpbf}(c).

It should be noted that the DARPA study collected samples from multiple
other sources, and that there is published, publicly available gene
expression data from the peripheral blood of the same patients we have
examined here. That data is analyzed in \citet{zaas09} and a time course
trajectory model is developed on a more complete version of the data in
\citet{carin11}. Together with the proteomics data included in
the supplementary material [\citet{Henetal13N2}], these offer interesting possibilities for future work into
jointly modeling proteomics and gene expression data. We have briefly
examined the correlation between protein and matched gene expression in
these data sets, but find that it is generally quite low. However, an
examination of the top genes discovered in \citet{zaas09} and the five
discriminative proteins elucidated here shows a high overlap in
associated pathways. We suspect that a comprehensive joint analysis of
these data is complicated by the tissue of origin. Specifically, it is
not clear that the proteins in blood plasma originate from peripheral
blood mononuclear cells (from which there is published gene expression
data). Instead, it is likely that much of the observed protein
expression is due to activities in organs such as the liver or kidneys
and from the endothelial lining of blood vessels.\looseness=1
%
\section{Concluding remarks}
We have presented a factor model specifically designed for proteomics
data analysis. It successfully handles broad scale variability that is
known to come from technical sources (such as batch effects and isotope
group specific noise), hence enabling us to estimate latent protein
profiles that better describe biological variability. Our hierarchical
representation of isotope groups, latent proteins and protein pathways
provides us with detailed annotation uncertainty assessment, detection
of possibly inaccurately annotated isotope groups and clustering of
proteins with similar expression profiles that reflect biologically
related interactions. We have also shown that features of our model can
be used to define predictive models based either on latent proteins or
groups of latent proteins.

\begin{appendix}\label{app}
\section*{Appendix: MCMC inference details}
We describe next the MCMC analysis mostly based on Gibbs sampling. We
provide then the relevant conditional posteriors and SMC-based update
details for the tree structure. To simplify notation, we use the
following shorthands. Let $\X^m=[x_1^m \en\cdots\en x_{N_m}^m]$ and $\X
=[\X^1 \en\cdots\en \X^{N_B}]$, where $N_B$ is the number of batches,
$N_m$ is the number of samples in batch $m$ and $N=\sum
_{m=1}^{N_B}N_m$. Define $\one_k$ to be a $k$-dimensional row vector of
ones and let $\widetilde{\X}$ be the full data set with the appropriate
means subtracted off; this is $\widetilde{\X}=[\X^1-\bmu^1 \one_{N_1}
\en
\cdots\en\X^{N_B}-\bmu^{N_B} \one_{K_{N_B}}]$, and $\Z=[{\mathbf
z}_1 \en
\cdots\en{\mathbf z}
_{N}]$ and $\W=[{\mathbf w}_1 \en\cdots\en{\mathbf w}_{N}]$,
systematic factors
and latent
protein matrices of sizes $N_F\times N$ and $N_P\times N$,
respectively. For any matrix $\M$, define $\M_{i:}$ as its $i$th row
and $\M_{:j}$ to be its $j$th column.
\subsection*{Noise variance} Sample each element of the diagonal of
$\bPsi
$ from
\[
\psi_i^{-1}|t_s,t_r \sim\Ga
\biggl(t_s+\frac{N}{2},t_r+c \biggr),
\]
where $t_s$ and $t_r$ are, respectively, prior shape and rate and
\[
c = \tfrac{1}{2}(\widetilde{\X}_{i:}-\A_{i:}\Z-
\B_{i:}\W) (\widetilde{\X}_{i:}-\A_{i:}\Z-
\B_{i:}\W)^\top.
\]
\subsection*{Batch means} Sample mean vector for batch $m$ from
\[
\bmu^m|t_m,t_p \sim\DN\Biggl(\C\middle
\Biggl(t_mt_p+\bPsi^{-1}\sum
_{n=1}^{N_m} {\mathbf x}_n^m-
\A{\mathbf z}_n-\B{\mathbf w}_n\middle\Biggr),\C
\Biggr),
\]
where $\C=(t_p+N_m\bPsi^{-1})^{-1}$, $t_m$ and $t_p$ are prior mean
and precision.
\subsection*{Systematic effect factors} The conditional posterior of
$\Z
$, using a scale mixture of Gaussian representation, can be computed
independently for each element of the matrix using
\[
z_{jn}|\tau_{jn} \sim\DN\bigl(c_{jn}
\A_{:j}^\top\bPsi^{-1} \bepsilon_{\setminus jn},c_{jn}
\bigr),
\]
where $c_{jn} = (\A_{:j}^\top\bPsi^{-1}\A_{:j}+\tau
_{jn}^{-1})^{-1}$ and
$\bepsilon_{\setminus jn}={\mathbf x}_{n}-\A{\mathbf z}_{n}-\B
{\mathbf
w}_n-\bmu^m| z_{jn}=0$.
The mixing variances $\tau_{jn}$ are exponentially distributed with
rate $\lambda^2$, hence, their resulting conditional posterior is
\[
\tau_{jn}^{-1}|\lambda^2 \sim\IG\biggl(\sqrt{
\frac{\lambda
^2}{z_{jn}}},\lambda^2 \biggr),\qquad \lambda^2|
\ell_s,\ell_r \sim\Ga\biggl(\ell_s+
\frac{1}{2},\ell_r+\frac{1}{2}\sum
_{j,n}\tau_{jn} \biggr),
\]
where $\ell_s$ and $\ell_r$ are shape and rate priors, respectively.
$\IG(\cdot|\mu,\lambda)$ is the inverse Gaussian distribution with mean
$\mu$ and scale $\lambda$ [\citet{chhikara89}]. Each element $a_{ij}$
from the loading matrix $\A$ is sampled from
\[
a_{ij} \sim\DN\bigl(c_{ij}\bepsilon_{\setminus ij}\Z
_{l:}^\top,c_{ij}\psi_i \bigr),
\]
where $c_{ij} = (\Z_{j:}\Z_{j:}^\top+\psi_{i}\rho_j)^{-1}$ and
$\bepsilon
_{\setminus ij}=\widetilde{\X}_{i:}-\A_{i:}\Z-\B_{i:}\W|a_{ij}=0$.
Then, column-wise precisions for $\A$ are drawn from
\[
\rho_j|r_s,r_r \sim\Ga
\biggl(r_s+\frac{p}{2},r_r+\sum
_i a_{ij}^2 \biggr),
\]
where $r_s$ and $r_r$ are prior shape and rate, respectively.
\subsection*{Protein profiles} The conditional posterior for latent
proteins $\W$ can be updated from
\[
\W_{k:}|\v_k \sim\DN\bigl(\C\B_{:k}^\top
\bPsi^{-1} (\widetilde{\X}-\A\Z)+\C\S_k^{-1}{
\mathbf m}_k,\C\bigr),
\]
where\vspace*{1pt} $\C=(\B_{:k}^\top\bPsi^{-1}\B_{:k}+\S_k^{-1})^{-1}$, with
${\mathbf m}_k$ and
$\S_k$ being mean and covariance of the parent profile $\v_k$ of $\W
_{k:}$. Note that $b_{ik}=0$ for all isotope groups not assumed to be
part of this protein, and that these will not contribute to the update
distribution for $\W_{k:}$. Besides,
\[
b_{ik}|b_{ik}\neq0 \sim\DN_+ \bigl(c(\widetilde{
\X}_{i:}-\A_{i:}\Z)\W_{k:}^\top,c
\psi_i \bigr),
\]
where $c = (\W_{k:}\W_{k:}^\top+\psi_i)^{-1}$ and $\DN_+(\cdot)$
is the
Gaussian distribution truncated below zero. Now we can sample IG-latent
protein assignments from
\begin{eqnarray*}
&\displaystyle u_i|\alpha,\bkappa,t_s,t_r \sim
\Dis(\v_i),&
\\
&\displaystyle v_{k} \propto (\alpha+n_k)c^{-{1}/{2}}
\bigl(t_r + \tfrac{1}{2} \widetilde{\X}_{i:}
\widetilde{\X}_{i:}^\top- \tfrac{1}{2}c^{-1}
\widetilde{\X}_{i:}\W_{k:}^\top\W_{k:}
\widetilde{\X}_{i:}^\top\bigr)^{-
(t_s+
{N}/{2} )},&
\end{eqnarray*}
where $n_k$ is the number of nonzero entries in column $k$ of $\B$,
$c=\W_{k:}\W_{k:}^\top$ and $v_{k}$ is the $k$th element of $\v_i$.

\subsection*{Protein pathway expression and tree structure} We sample the
tree structure components $\t$, $\bpi$ and $\bPhi$, and the means and
covariances of each internal node of the tree, ${\mathbf m}_k$ and $\S_k$,
respectively, using the SMC sampler described in \citet{henao12b}. In
particular, $\{\t,\bpi\}$ are obtained for a number $M$ of particles,
as a leaves to root SMC pass, together with partial updates of the node
parameters $\{{\mathbf m}_k,\S_k\}$. Next we use the particle's
weights to
sample a single configuration. The procedure is completed by resampling
the hyperparameters of the covariance function and by completing the
updates of the node parameters using the selected configuration, the
latter as a root to leaves pass.
\subsection*{Missing values} For each missing value $x_{in}^m$
corresponding to isotope group~$i$, sample $n$ and batch $m$, we simply
use independent standardized Gaussian prior distributions.
\subsection*{Initialization} We start the model from maximum likelihood
estimates of the less critical quantities, that is, batch means $\{\bmu
^m\}_{m=1}^{N_B}$ and noise variances $\bPsi$. Systematic factors $\Z$
and latent proteins $\W$ are initialized using standardized Gaussian
distributions. The loading matrices $\A$ and $\B$ (nonzero elements
only) were set to ordinary least squares estimates based upon already
set $\Z$ and $\W$, respectively. The vector of associations $\u$ was
set with the information obtained from annotation about IG-protein assignments.
\end{appendix}

\section*{Acknowledgments}

We thank the Editor and the anonymous referees for their helpful
comments and discussions that improved the presentation of this paper.

\begin{supplement}
\stitle{Tree structure}
\slink[doi]{10.1214/13-AOAS639SUPPA} 
\sdatatype{.eps}
\sfilename{aoas639\_suppa.eps}
\sdescription{Figure showing the tree structure for the H1N1/H3N2
viral challenge data.}
\end{supplement}

\begin{supplement}
\stitle{Data}
\slink[doi]{10.1214/13-AOAS639SUPPB} 
\sdatatype{.zip}
\sfilename{aoas639\_suppb.zip}
\sdescription{H1N1/H3N2 viral challenge raw data.}
\end{supplement}

\begin{supplement}
\stitle{Estimated proteins}
\slink[doi]{10.1214/13-AOAS639SUPPC} 
\sdatatype{.pdf}
\sfilename{aoas639\_suppc.pdf}
\sdescription{Figures showing the estimated proteins for the spike-in
data experiment.}
\end{supplement}

%

\printaddresses

\end{document}